\documentclass[12pt]{iopart}

\begin{document}

\title[Gravitational quantum states]{Gravitational quantum states as finite representations of the Lorentz group}

\author{Francesco Cianfrani}

\address{CNRS, Aix-Marseille Universite, PIIM UMR7345, Avenue Escadrille Normandie Niemen, 13013 Marseille, France.}
\ead{francesco.cianfrani@univ-amu.fr}
\begin{abstract}
A manifestly Lorentz-covariant formulation of Loop Quantum Gravity (LQG) is given in terms of finite-dimensional representations of the Lorentz group. The formulation accounts for discrete symmetries, such as parity and time-reversal, and it establishes a link with Wigner classification of particles. The resulting quantum model can be seen as LQG-like with the internal $SU(2)\otimes SU(2)$ group and it is free of the Immirzi parameter, while the scalar constraint is just the Euclidean part.  
\end{abstract}


\maketitle

\section{Introduction} 
A quantum theory of General Relativity is expected to provide a quantum description of the geometry and of its interaction with fundamental particle fields. Wigner classification \cite{Wigner:39,Bargmann:48} describes particles in terms of finite-dimensional representations of the Lorentz group, which are developed as irreducible representations of two commuting SU(2) groups (see also \cite{Weinberg:02}). The action of Lorentz transformations on such representations is obtained by a complexification of each ${\it su}(2)$ algebra. Mathematically, the representations of a (noncompact) group, the Lorentz group, are constructed from those of a (compact) subgroup, $SU(2)\otimes SU(2)$.  

Loop Quantum Gravity (LQG) (see \cite{revloop,Thiemannb} for some reviews) provides a quantum description of the geometry in terms of an internal SU(2) group, that can be identified with the group of rotations. The SU(2)-invariant model can be derived from the Lorentz invariant vier-bein formulation of gravity by a partial gauge-fixing that has also the merit of removing some second-class constraints \cite{Cianfrani:09}. In order to get the Gauss constraint of SU(2) Yang-Mills theory, a canonical transformation must be performed in phase space and it introduces a fundamental ambiguity, the Immirzi parameter, that affect the spectra of quantum geometrical operators \cite{Immirzi}. On a quantum level, the holonomy-flux algebra is quantized and the corresponding Hilbert space is defined on a graph as the direct product of irreducible $SU(2)$ representations along all the edges of the graph (spin-network functions). The measure is inherited from the SU(2) Haar measure at each edge and since SU(2) is compact it is positive defined \cite{ALMMT95,LOST06}. The resulting picture of the quantum geometry is that the space dual to the graph is endowed with discretized volumes and areas \cite{discr}. Nevertheless, the full quantization program has not been completed due to the technical difficulties in implementing the scalar constraint (Hamiltonian) operator \cite{Thiemann:98qsd} and the proper continuum limit \cite{Thiemann:07}.     

In this work, loop quantization of the Einstein-Hilbert action is performed in a manifestly Lorentz-covariant formulation. In the same spirit of Wigner classification, Lorentz group representations are constructed as $SU(2)\otimes SU(2)$ irreducible representations and the measure, inherited from the corresponding Haar measure, is positive defined. It will be shown that the resulting spin-network functions are invariant under proper orthochronus transformations, that can be realized by complexification of the two ${\it su}(2)$ algebras. Furthermore, since parity and time-reversal operators exchange the two $SU(2)$ group elements among themselves, fully \emph{Lorentz-invariant} states can be defined by a symmetrization that is the analogous of Wigner symmetrization of representations $(j_1,j_2)=(j_1,j_2)\oplus (j_2,j_1)$ (balanced representations).    

The key-point of this analysis is a reformulation of gravity in terms of Lorentz spin connections and corresponding momenta, which is free of second-class constraints. Besides the standard constraints implementing diffeomorphisms and local Lorentz invariance, the conditions $C^{ab}=\epsilon^{IJKL}\, \pi^a_{IJ}\, \pi^b_{KL} =0$ are present and the corresponding operators annihilate those states constructed with balanced representations. 
 
Hence, the obtained formulation is the analogous of standard LQG modulo the replacement of $SU(2)$ with $SU(2)\otimes SU(2)$ and has the following nice features
\begin{itemize} 
\item it is free of Immirzi ambiguity,
\item the scalar constraint is just the so-called Euclidean term.
\end{itemize}
Quantum geometrical operators are peculiar: the area is the sum of the two LQG areas along each $SU(2)$ subgroup, the volume is the difference of the analogous LQG operators.  

This model can be seen as the quantization of the original Ashtekar self-dual and anti-self dual connections \cite{Ashtekar:86}, with the prescription of considering both of them. The reality condition on momenta, that plagued the original Ashtekar proposal, is here implemented through the reality of the algebra representation, that is insured by parity invariance through the symmetrization of the $SU(2)\otimes SU(2)$ representations. 

It is worth noting the similarity between the presented framework and Barret-Crane model for Euclidean Quantum Gravity \cite{Barrett:98}, in which the same kind of representations are derived from the universal covering of the SO(4) group. However, the extension of the original Barret-Crane model to Lorentzian gravity led to spin-foam models \cite{Barrett:00,Crane:01}, that has been constructed from the infinite-dimensional unitary representations of the Lorentz group, thus losing contact with Wigner representations.

Instead, the description of quantum gravitational degrees of freedom in terms of Lorentz finite representation makes this theory the natural arena for investigating the interaction between quantum geometry and fundamental fields. Although matter fields have already been considered in LQG \cite{Thiemann:98}, here for the first time one is accounting for the relativistic symmetries of quantum gravitational states, included parity and time-reversal, that could open a novel perspective on the interplay with Quantum Field Theory.  

The manuscript is organized as follows. In section \ref{sec0} it is discussed how the standard formulation of LQG is based on a partial gauge fixing of the symmetry under local Lorentz transformations. In section \ref{sec1} the covariant Hamiltonian formulation of Einstein-Hilbert action is presented and the system of constraints is shown to be first-class. In section \ref{sec2} the LQG quantization scheme is reviewed and in section \ref{sec3} it is outlined how Lorentz invariant state can be constructed. These states are shown to be annihilated by the quantum constraint corresponding to $C^{ab}=0$ in section \ref{sec4}, while in section \ref{sec5} and in section \ref{sec6} the action of geometrical operators and the scalar constraint operator are discussed, respectively. Brief concluding remarks follow in section \ref{sec7}.

\section{Partial gauge fixing in LQG}\label{sec0}
The classical formulation of LQG is based on a parametrization of the phase space in terms of Ashtekar-Barbero connections $A^i_a$ and  
inverse densitized triads $E_i^a$
\begin{equation}
A^i_a=\omega_a^{0i}-\frac{1}{2\gamma}\epsilon^i_{\phantom1jk}\,\omega^{jk}_a\,,
\end{equation}
$\omega^{IJ}_a$ being Lorentz spin connections, while $\gamma$ is the Immirzi parameter. 

The classical Hamiltonian can be derived from the standard ADM formulation of gravity by writing the 3-metric $h_{ab}=\delta_{ij}\,e^i_a\,e^j_b$ in terms of triads $e^a_i$, which introduces an additional gauge symmetry, since the metric is determined up to an internal rotation. In phase space a gauge symmetry is associated to a first class constraint. Through a proper canonical transformation labeled by the Immirzi parameter \cite{Barbero:95}, the constraint associated to the internal rotational symmetry coincides with the Gauss constraint of a $SU(2)$ Yang-Mills gauge theory. This achievement is crucial, since it allows us to define $SU(2)$ holonomies and to parametrize the classical phase space in terms of them and of the corresponding fluxes (see section \ref{sec2}). 

An alternative derivation of the LQG Hamiltonian can be given starting from the Einstein-Hilbert action plus the so-called Holst term \cite{Ho96}, which in vier-bein formulation reads        
\begin{equation} 
S_{LQG}=\frac{1}{2}\int e\,\left(e^\mu_I\, e^\nu_I-\frac{1}{2\gamma}\epsilon_{IJ}^{\phantom{12}KL} e^\mu_K\, e^\nu_L\right) \,R_{\mu\nu}^{IJ}(\omega)\,d^4x,
\end{equation}
where $e^I_\mu$ denotes the vier-bein of the space-time metric $g_{\mu\nu}=\eta_{IJ}\,e^I_\mu \,e^J_\nu$, $\omega^{IJ}_\mu$ are the spin-connections and the curvature 2-form $R^{IJ}$ reads explicitly
\begin{equation} 
R^{IJ}= d\omega^{IJ}+ \omega^I_{\phantom{} K}\wedge \omega^{KJ} \,.\label{omega}
\end{equation}
and $e$ denotes the determinant of the 4x4 matrix $e^I_\mu$. In order to obtain LQG phase space the vier-bein vectors $e^I_\mu$ must be written as
\begin{equation}  
e^I_\mu=\left(\begin{array}{cc} N & N^a\,e^i_a \\ 0 & e^i_a \end{array}\right)\,,\label{etg}
\end{equation}
$N$ and $N^a$ being the lapse function and the shift vector of the corresponding ADM formulation, respectively. In particular, the expression (\ref{etg}) implies the condition $e^0_a=0$, which is a partial gauge fixing of the symmetry under local Lorentz transformations  known as the time-gauge condition. It implies that the local Lorentz frame is adapted to the 3+1 slicing of the space-time manifold so that the vier-bein components $e^i_a$ coincides with the triads of the 3-metric (which is not necessarily the case in a generic local Lorentz frame). 

Therefore, the classical formulation of LQG is based on a partial gauge fixing of the symmetry under local Lorentz transformations. While classically it is not an issue to fix some symmetries, on a quantum level it matters if a symmetry is fixed before quantization or extended to the quantum Hilbert space. For instance, different inequivalent approaches to the problem of time have been developed in Quantum Gravity in the two cases. 

A classical formulation with a generic local Lorentz frame has been proposed in \cite{BS01} by removing the time-gauge condition 
\begin{equation}  
e^I_\mu=\left(\begin{array}{cc} N & N^a\,e^i_a \\ \chi_i\,e^i_a & e^i_a \end{array}\right)\,,
\end{equation}
where the functions $\chi_i$ are the boost parameters of the local Lorentz frame with respect to the 3+1 slicing of the space-time manifold. 
It has been outlined in \cite{BS01} how the whole system of constraints is second-class in a generic local Lorentz frame and it has been proposed a classical solution reducing the constraints to be first-class in a generic fixed frame. 

The second-class character of the system of constraints provides a strong complication to a covariant quantum formulation. In fact, if second-class constraint are present one has two options: i) to solve some constraints classically in order to get a first-class system, which implies loosing covariance (as in \cite{Cianfrani:09}), ii) to work with the original unconstrained phase-space coordinate by replacing Poisson brackets with Dirac ones, which provides a much more complex algebra between connections and momenta, for which no quantum representation is usually known (see for instance the algebra in \cite{Alexandrov:2006}).

In what follows, a covariant formulation free of second-class constraints is proposed. 

\section{Hamiltonian formulation} \label{sec1}
The Einstein-Hilbert action is here considered, {\it i.e.} 
\begin{equation} 
S_{EH}=\frac{1}{2}\int e\,e^\mu_I\, e^\nu_I \,R_{\mu\nu}^{IJ}(\omega)\,d^4x.\label{act}
\end{equation}
and the corresponding Hamiltonian formulation (see \cite{Cianfrani:09} in the limit $\gamma\rightarrow \infty$) is performed by taking spin connections $\omega^{IJ}_a$ as configuration variables, whose corresponding momenta are given in terms of inverse vierbein vectors as
\begin{equation}    
\pi^a_{IJ}=\frac{\delta S_{EH}}{\delta \partial_t\omega^{IJ}_a}= e\, e^{t}_{[I}\, e^a_{J]}\,.\label{momenta}
\end{equation}

By direct substitution, one can verify that the following conditions hold 
\begin{equation}    
C^{ab}= \epsilon^{IJKL}\pi_{IJ}^{(a}\pi_{KL}^{b)}=0 \,.\label{Cab}
\end{equation} 

By performing a Lagrange transformation, the total hamiltonian is a linear combination of primary constraints 
\begin{eqnarray}   
\mathcal{H}=\int\bigg[\frac{1}{\sqrt{-g^{tt}}}\,\mathcal{S}-\frac{g^{ta}}{g^{tt}}\,\mathcal{V}_a-\omega^{IJ}_t\,G_{IJ}+\lambda_{ab}C^{ab}+\lambda^{IJ}\pi_{IJ}^t\bigg]d^3x,
\end{eqnarray}
where $g^{tt}$, $g^{ta}$, $\omega^{IJ}_t$, $\lambda_{ab}$ and $\lambda^{IJ}$ behave as Lagrangian multipliers. 

In particular, in the subspace $\{\omega^{IJ}_a,\pi^a_{IJ}\}$ the constraint hypersurfaces is parametrized by the conditions
\begin{equation}   
\left\{\begin{array}{c}\mathcal{S}=|\pi|^{-1/2}\,\pi^a_{IK}\,\pi^{bK}_{\phantom1J}\,R^{IJ}_{ab}=0 \\\\ 
\mathcal{V}_a=\pi^b_{IJ}\,R^{IJ}_{ab}=0 \\\\ G_{IJ}=D_a\pi^a_{IJ}=\partial_a\pi^a_{IJ}-\omega_{[I\phantom2a}^{\phantom1K}\pi^a_{|K|J]}=0 \\\\ C^{ab}=\epsilon^{IJKL}\pi_{IJ}^{(a}\pi_{KL}^{b)}=0 
\end{array}\right.,\label{const}
\end{equation}
where the metric determinant $\pi$ reads 
\begin{equation}
\pi=\frac{1}{3!} \,\epsilon_{abc}\,\epsilon^{IJKL} \pi^{a\phantom1 M}_I\,\pi^b_{MJ}\,\pi^c_{KL}\,.\label{pi}
\end{equation}

One could wonder whether there are more primary constraints. The answer is negative. In fact, the above constraints exhaust all the known Lagrangian symmetries. The vanishing of the scalar $\mathcal{S}$ and vector $\mathcal{V}_i$ constraints is due to the invariance under time shifts $t\rightarrow t+f(t)$ and space-like diffeomorphisms $x\rightarrow x^i+ \xi^i(x)$, respectively, while $G_{IJ}$ is the Gauss constraint of the local Lorentz symmetry (it generates Lorentz tranformations in the tangent space). 
 
The condition $C^{ab}=0$ is due to the fact that there are less independent components on the right-hand sides of Eq.(\ref{momenta}) with respect to the total number of momenta. The momenta $\pi^a_{IJ}$ are in all $6\times 3=18$, but the right-hand side contains $4\times 3=12$ independent components $e^a_{J}$\footnote{One should not count $e^t_I$ since they enter the definition of the Lagrange multipliers $\frac{1}{\sqrt{-g^{tt}}}$ and $\frac{g^{ta}}{g^{tt}}$, so they are Lagrangian multipliers themselves.}, thus they are not completely independent. Hence, $6$ conditions are needed and they are precisely $C^{ab}=0$.

The secondary constraints can now be computed by performing Poisson brackets among the primary ones. It results that they all vanish on the constraint hypersurface (\ref{const}) and the system of constraints is first-class.

In previous analysis \cite{Cianfrani:09,Cianfrani:14} it has been reported that the total system of constraints is second-class, because  the Poisson brackets $\{C^{ab},\mathcal{S}\}$ do not vanish and they read explicitly
\begin{equation}
D^{ab}=|\pi|^{-1/2}\epsilon^{IJKL}\pi^c_{IM}\pi^{(aM}_{\phantom1\phantom2J}D_c\pi^{b)}_{KL}\label{D}\,.
\end{equation}

The novel result of this work is that indeed $D^{ab}$ vanishes identically. By using the symmetry under the exchange of spatial indexes $a, b$ and by moving the derivative $D_c$ it can be shown that\footnote{It is useful the relation 
\begin{equation}
D^{ab}=|\pi|^{-1/2}\epsilon^{IJKL}\pi^c_{IM}\pi^{(b}_{KL}D_c \pi^{a)M}_{\phantom1\phantom2J}\,,
\end{equation}
 which can be demonstrated using the formulas for the product of two $\epsilon$ skew-symmetric tensors.}
\begin{eqnarray}
D^{ab}=&\frac{1}{2}\,|\pi|^{-1/2}\partial_c\left(\epsilon^{IJKL}\pi^c_{IM}\pi^{(aM}_{\phantom1\phantom2J}\pi^{b)}_{KL}\right)-\frac{1}{2}\,|\pi|^{-1/2}\epsilon^{IJKL}D_c\pi^c_{IM}\pi^{(aM}_{\phantom1\phantom2J}\pi^{b)}_{KL}\,.\nonumber
\end{eqnarray} 

The second term vanishes because of local Lorentz invariance ($D_c\pi^c_{IM}= G_{IM}=0$), while the first term vanishes since 
\begin{equation}
\epsilon^{IJKL}\,\pi^{(a}_{MJ}\pi^{b)}_{KL} =\frac{1}{4}\, \delta^I_M \,\epsilon^{NJKL}\,\pi^{(a}_{NJ}\pi^{b)}_{KL} = \delta^I_M\, C^{ab} =0\,.
\end{equation} 

Therefore, the system of constraints is first-class for Einstein-Hilbert gravity with spin connections as configuration variables. This also implies that all the constraints can be implemented as operators annhilating physical states on a quantum level.

\section{Canonical quantization and LQG} \label{sec2}
The canonical quantization program is based on representing in a Hilbert space the Hamiltonian constraints as operators that annihilates the physical quantum states. LQG succeed in the definition of the so-called kinematical Hilbert space, {\it i.e.} an Hilbert space in which the quantum states that are invariant under the action of an internal compact gauge group (SU(2) in standard LQG, but more general compact groups can be considered \cite{Bodendorfer:13}) and of space-like diffeomorphisms can be defined. The kinematical Hilbert space is defined from the space $Cyl$ of cylindrical functions, {\it i.e.} the space of continuous functions of holonomies along the edges of a graph. The holonomies are constructed as path-ordered exponentials of the connections and they are elements of the internal group, SU(2), while the momenta smeared over spatial surfaces acts as derivative operators that provide the insertion of the algebra generator times a factor $\pm 1$ depending on the relative orientation between the edge and the surface. The measure in $Cyl$ is defined as the product over all the edges of the SU(2) Haar measure and a basis is given by spin-network functions, that are obtained by expanding group elements $g$ in irreducible SU(2) representations $D^j_{m_en_e}(g)$, labeled by the spin number $j_e$ and the magnetic numbers $m_e$ and $n_e$\footnote{The two magnetic numbers $m_e$ and $n_e$ corresponds to the start and end point of the edge}. 

The internal SU(2) gauge symmetry is implemented by inserting invariant intertwiners at the nodes of the graph: since holonomies transform by the insertion of SU(2) group elements at the start and end points of the edges, gauge invariant states can be constructed by connecting edges through invariant tensors ${\it i}_{m_1m_2..}^{m_k m_{k+1}..}$ at nodes\footnote{\label{1}Here subscript/superscript are magnetic indexes of outgoing/incoming edges at the node.}
\begin{equation}    
{\it i}_{m_{1}m_{2}}^{m_{k}m_{k+1}..}\,G_{m'_{1}}^{m_{1}}\,G_{m'_{2}}^{m_{2}}\,...\,(G^{-1})_{m_{k}}^{m'_{k}}\,(G^{-1})_{m_{k+1}}^{m'_{k+1}}...\,=\,{\it i}_{m'_{1}m'_{2}}^{m'_{k}m'_{k+1}..}\,.
\end{equation}
Such invariant tensors are derived in SU(2) recoupling theory \cite{Brink:68,Makinen:19} and for a generic $n$-valent node they can be derived by contracting each couple of edges through the fundamental three-valent intertwiner ${\it i}_{m_1m_2}^{m_3}$ that is the Clebsch-Gordan coefficient for the expansion of the spin state  $|j_3,m_3\rangle$ into $|j_1,m_1\rangle \otimes |j_2,m_2\rangle$.   

The invariance under space-like diffeomorphisms is formally implemented by considering states defined over s-knots \cite{LOST06}, {\it i.e.} over the equivalence class of diffeomorphsims-related graphs, such that only the topological properties of the graph are relevant.    

Therefore, a state of the kinematical Hilbert space is a linear combination of spin-network functions $\Psi=\sum_s c_s\,\Psi_s$ that are  labeled by the collections $\{j_e\}$ of spin numbers at edges and of invariant interwiners $\{{\it i}_v\}$ at nodes:
\begin{equation}
\Psi_s(\{g\})= \prod_v {\it i}_v \,\prod_e \,D^{j_e}(g_e)\,,
\end{equation}
where the magnetic indexes of ${\it i}_v$ are properly contracted with the magnetic indexes of $D^{j_e}(g_e)$ for all the edges $e$ emanating from $v$.

The remaining constraint $\mathcal{S}$ can be represented as a self-adjoint operator, but its expression is very complicated and some quantum ambiguities remain so that the canonical quantization program of LQG has not been able to provide an explicit expression for physical quantum states.

\section{Lorentz internal symmetry} \label{sec3}

It is well known that the Lorentz algebra ${\it so}(1,3)$ is isomorphic to the direct sum of two complex-conjugate ${\it sl}(2,C)$ algebra and that each ${\it sl}(2,C)$ can be seen as the complexification of the ${\it su}(2)$ algebra. The two ${\it su}(2)$ generators $J^{A}_i$ and $J^B_i$ are related to the generators of rotations $R_i=\epsilon_{ijk}J_{jk}$ and boosts $K_i=J_{0i}$ as follows
\begin{equation}      
J^{A}_i=\displaystyle\frac{R_i+iK_i}{2}\qquad J^{B}_i=\displaystyle\frac{R_i-iK_i}{2}\,.\label{RKJ}
\end{equation}
Finite dimensional representations of the orthochronous Lorentz group can be constructed as the tensor product of the two SU(2) representations $(j_1 , j_2)=|j_1,m_1\rangle_A \otimes |j_2,m_2\rangle_B$ and the action of a generic element of SO(1,3)$^+$ is given by
\begin{eqnarray}
\label{lorentz+}
|j_1,m_1\rangle_A \otimes |j_2,m_2\rangle_B & \rightarrow e^{i\frac{\vec{\theta}\cdot \vec{J}^A}{2}-\frac{\vec{\eta}\cdot \vec{J}^A}{2}} &|j_1,m_1\rangle_A \otimes e^{i\frac{\vec{\theta}\cdot \vec{J}^B}{2}+\frac{\vec{\eta}\cdot \vec{J}^B}{2}}\, |j_2,m_2\rangle_B\,,
\end{eqnarray}
$\theta^i$ and $\eta^i$ being rotation angles and boosts, respectively. Parity and time reversal exchange $J^A_i$ with $J^B_i$ \footnote{The representation of $T$ is linear and unitary.}:
\begin{eqnarray}
P\,|j_1,m_1\rangle_A \otimes &|j_2,m_2\rangle_B = |j_2,m_2\rangle_A \otimes |j_1,m_1\rangle_B\\
T \,|j_1,m_1\rangle_A \otimes &|j_2,m_2\rangle_B =\nonumber\\ &(-1)^{j_2}\hat{\varepsilon}_{m_2,m'_2}|j_2,m'_2\rangle_A \otimes (-1)^{j_1}\hat{\varepsilon}_{m_1,m'_1}|j_1,m'_1\rangle_B\,, 
\end{eqnarray}
$\hat{\varepsilon}_{m,m'}=(-1)^{j-m} \delta_{m,-m'}$ being the operator raising and lowering magnetic indexes.
 
Hence, irreducible representations of the full Lorentz group are obtain by performing the direct sum $(j_1, j_2) \oplus (j_2, j_1)$. The most relevant among such kind of representations are Dirac 4-spinors $(1/2, 0) \oplus (0, 1/2)$. The representation of the Lorentz algebra  on $(j_1, j_2) \oplus (j_2, j_1)$ is given by matrices with real elements and since a generic element rewrites as
\begin{equation}
\theta^iR_i+\eta^iK_i= (\theta^i+i\eta_i)J^A_i+(\theta^i-i\eta_i)J^B_i\,,  
\end{equation}
such reality condition implies that the two ${\it SU}(2)$ generators $J^A_i$ and $J^B_i$ are complex conjugate. 

The classical phase space is here describe by the holonomies of the Lorentz group $h_e(\omega)$ and the corresponding smeared fluxes $\pi_{IJ}(S)=\int \pi_{IJ}^a\,dS_a$ such that the holonomy-flux algebra reads 
\begin{eqnarray}
\{\pi_{IJ}(S), h_e(\omega)\}= i\,J_{IJ}\, h_e(\omega)\,,\label{hf-alg}
 \end{eqnarray}
if the edge $e$ is dual to the surface $S$ (for simplicity it is also assumed outgoing), otherwise it vanishes.

The kinematical Hilbert space can be defined by the quantization of the space of cylindrical functions over Lorentz holonomies and by promoting smeared momenta to operators by a representation of the holonomy-flux algebra (\ref{hf-alg}), {\it i.e.} 
\begin{eqnarray}
\hat{\pi}_{IJ}(S)\, h_e(\omega)= i\,J_{IJ}\, h_e(\omega)\,.\label{hf-alg-q}
 \end{eqnarray}

The measure inherited from the Haar measure of Lorentz group is not positive definite, since the group is noncompact, and thus it is not suitable for the definition of the quantum scalar product. This is the reason why in LQG the group is assumed to be compact (see chapter 6 in \cite{Thiemannb}).  

The main idea of this work is to expand the Lorentz group elements at edges in terms of the non-unitary finite irreducible representations of the Lorentz group, namely the irreducible representations of $SU(2)_A \otimes SU(2)_B$, and to inherit the measure from the Haar measure of $SU(2)_A \otimes SU(2)_B$, which is positive definite.  

Hence, spin-network states are defined in terms of the direct product of two $SU(2)$ representations at each edge $D^{j^{(A)}_e}(g^{(A)}_e) \otimes D^{j^{(B)}_e}(g^{(B)}_e)$, $g^{(A)}_e$ and $g^{(B)}_e$ being the $SU(2)$ elements corresponding to the self-dual and antiself-dual parts of the Lorentz algebra, while $D^{j_e}$ denotes the analytic continuation of Wigner matrix in the $j_e$ representation. For each Wigner matrix the measure reads
\begin{equation}
\langle D^{j'_e}_{m'_en'_e} | D^{j_e}_{m_en_e} \rangle = \delta_{j_ej'_e}\,\delta_{m_em'_e}\delta_{n_en'_e}\,. 
\end{equation}

In order words, the kinematical Hilbert space is constructed from the space of cylindrical functions over ${\it so}(1,3)$ connection, basis elements are constructed by an expansion in terms of finite $SU(2)_A \otimes SU(2)_B$ representations and the measure is defined as the $SU(2)_A \otimes SU(2)_B$ Haar measure.

Furthermore, in order to construct Lorentz invariant states the $SU(2)_A \otimes SU(2)_B$ invariant intertwiners ${\it i}^{(A)}_v \otimes {\it i}^{(B)}_v$ are inserted at nodes $v$, so getting the following states
\begin{eqnarray}
\Psi_s(\{g^{(A)}\},\{g^{(B)}\}) = \prod_v {\it i}^{(A)}_v \otimes {\it i}^{(B)}_v \,\prod_e \,D^{j^{(A)}_e}(g^{(A)}_e) \otimes D^{j^{(B)}_e}(g^{(B)}_e)\,,\label{spinnetAB}
\end{eqnarray}
where ${\it i}^{(A)}$ and ${\it i}^{(B)}$ are contracted with $D^{j^{(A)}_e}(g^{(A)}_e)$ and $D^{j^{(B)}_e}(g^{(B)}_e)$, respectively.

Proper orthochronous Lorentz transformations provides the insertion at the edge boundary points of those group elements in Eq.(\ref{lorentz+})
and the SU(2) interwiners ${\it i}^{(A)}$ and ${\it i}^{(B)}$ are invariant tensor with respect to them. In fact, a generic element $\Lambda\in SO(1,3)^+$ can be decomposed as $\Lambda= R(\theta) B R(\tilde\theta)$, where $R(\theta)$ and $R(\tilde\theta)$ are two rotations and $B$ is a boost along a given direction. It is worth noting that rotations act on each $SU(2)$ subgroup through the insertion of elements of the subgroup itself, thus the corresponding intertwiners are invariant under them by construction. Furthermore, the boost $B$ along the direction $3$ acts at each magnetic index through the matrix $e^{\mp\frac{J_3}{2}}=e^{\frac{m}{2}}\,\delta_{mm'}$ 
such that the invariance of the intertwiner can be proved as follows
\begin{eqnarray} 
{\it i}^{m_k ..}_{m_1m_2..}\,B_{m_1,m_1'}\,B_{m_2,m_2'}..\,B^{-1}_{m_k,m_k'}..=
{\it i}^{m'_k ..}_{m'_1m'_2..}\,e^{\mp \frac{m'_1}{2}}\,e^{\mp \frac{m'_2}{2}}..\,e^{\pm \frac{m'_k}{2}}={\it i}^{m'_k ..}_{m'_1m'_2..}\,,
\end{eqnarray}
where it has been used that $J^{..}_3 |j,m\rangle= m|j,m\rangle$ and the sums of incoming and outcoming magnetic numbers are equal (for a three-valent node ${\it i}^{m_3}_{m_1m_2}=C^{j_3m_3}_{j_1j_2m_1m_2}\propto \delta_{m_3,m_1+m_2}$). Hence, \emph{the spin-network states (\ref{spinnetAB}) are invariant under proper orthochronous Lorentz transformations}.

In order to construct Lorentz-invariant states, parity and time-reversal should be included. Parity exchange the two SU(2) group elements among each other while time-reversal exchange and reverse them, $D^j(g)\rightarrow D^j(g^{-1})$. Spin-network functions are already invariant under the reversal of the group elements at edges (this is due to the fact that the intertwiners do not change by lowering/raising all upper/lower indexes \cite{Makinen:19}).
Hence, taking the analogous of the $(j_1, j_2) \oplus (j_2, j_1)$ representation for spin-networks, one can define \emph{Lorentz invariant spin-networks} as
\begin{eqnarray}
\Psi_s(\{g^{(A)}\},\{g^{(B)}\})=& \prod_v {\it i}^{(A)}_v \otimes {\it i}^{(B)}_v \prod_e \,D^{j^{(A)}_e}(g^{(A)}_e) \otimes D^{j^{(B)}_e}(g^{(B)}_e)\nonumber\\
&+\prod_v {\it i}^{(B)}_v \otimes {\it i}^{(A)}_v  \prod_e \,D^{j^{(B)}_e}(g^{(A)}_e) \otimes D^{j^{(A)}_e}(g^{(B)}_e)\,.\label{LIsn}
\end{eqnarray}

It is worth noting the difference between Lorentz-invariant spin-networks above and projected spin-networks\cite{Alexandrov:03}, that have been define to embed SU(2) spin-networks into Lorentz ones. Projected spin-network have an additional label, given by the unit time-normal, and they are based on the decomposition in terms of infinite-dimensional unitary representations of the Lorentz group.  

\section{The constraint $C^{ab}=0$} \label{sec4}
The flux $\pi_{IJ}(S)$ of momenta across a surface $S$ can be defined as a quantum operator that provides the insertion of Lorentz algebra elements at the intersection points between $S$ and the edge $e$. One can always split the edges such that the intersection is at the starting point of $e$ and chose the positive surface orientation in the edge direction, so getting 
\begin{eqnarray}
\hat{\pi}_{IJ}(S) \,D^{j^{(A)}_e}(g^{(A)}_e) \otimes D^{j^{(B)}_e}&(g^{(B)}_e)=
 J^{(A)}_{IJ}D^{j^{(A)}_e}(g^{(A)}_e) \otimes D^{j^{(B)}_e}(g^{(B)}_e)\nonumber\\ 
&+ D^{j^{(A)}_e}(g^{(A)}_e) \otimes J^{(B)}_{IJ}D^{j^{(B)}_e}(g^{(B)}_e)\,,
\end{eqnarray}
Using the definitions of boost $K_i=J_{0i}$ and rotation $R_i=\frac{1}{2}\epsilon_{ijk}\,J_{jk}$ generators and Eqs.(\ref{RKJ}), one obtains 
\begin{eqnarray}
\hat{\pi}_{0i}(S) \,D^{j^{(A)}_e}(g^{(A)}_e) \otimes D^{j^{(B)}_e}&(g^{(B)}_e)=
 -i\,J^{(A)}_{i}D^{j^{(A)}_e}(g^{(A)}_e) \otimes D^{j^{(B)}_e}(g^{(B)}_e) \nonumber\\
& +i\,D^{j^{(A)}_e}(g^{(A)}_e) \otimes J^{(B)}_{i}D^{j^{(B)}_e}(g^{(B)}_e)\,,\label{pi0i+}
\end{eqnarray}
\begin{eqnarray}
\frac{1}{2}\epsilon_{ijk}\hat{\pi}_{jk}(S) \,D^{j^{(A)}_e}(g^{(A)}_e) \otimes D^{j^{(B)}_e}&(g^{(B)}_e)=
\,J^{(A)}_{i}D^{j^{(A)}_e}(g^{(A)}_e) \otimes D^{j^{(B)}_e}(g^{(B)}_e) 
\nonumber\\ &+\,D^{j^{(A)}_e}(g^{(A)}_e) \otimes J^{(B)}_{i}D^{j^{(B)}_e}(g^{(B)}_e)\,,\label{pijk+}
\end{eqnarray}

The constraint $C^{ab}(x)=0$ can be represented on a quantum level in terms of fluxes across two surfaces $S^a$ and $S^b$ centered around the point $x$ and whose normal vectors point in the directions $a$ and $b$, respectively. The only non trivial case when acting on a quantum state is when $x$ is a point of the graph on which the quantum state is based. If $x$ is not a node of the graph, one can focus on a single edge $e$ starting at $x$ and compute the action of momenta from Eqs.(\ref{pi0i+})-(\ref{pijk+}) so finding 
\begin{eqnarray}    
\epsilon^{IJKL} \pi_{IJ}(S^a)\,\pi_{KL}(S^b) \,D^{j^{(A)}_e}(g^{(A)}_e) \otimes D^{j^{(B)}_e}(g^{(B)}_e)=\\
-i\,J^{(A)}_iJ^{(A)}_iD^{j^{(A)}_e}(g^{(A)}_e) \otimes D^{j^{(B)}_e}(g^{(B)}_e) +i\,D^{j^{(A)}_e}(g^{(A)}_e) \otimes J^{(B)}_{i}J^{(B)}_{i}D^{j^{(B)}_e}(g^{(B)}_e)\,.\nonumber
\end{eqnarray}
Therefore, the action of the constraint does not mix up the different SU(2) group elements and it is skew-symmetric with respect to the exchange $j^{(A)}\rightarrow j^{(B)}$. This implies that it identically vanishes on those states, such as (\ref{LIsn}), that are symmetric for $j^{(A)}\rightarrow j^{(B)}$. The same conclusion holds if $x$ is a node of the graph, the only difference being that one must consider two edges instead of one. Therefore, \emph{Lorentz-invariant spin-network functions $\Psi_s$ are annihilated by the quantum constraint corresponding to $C^{ab}$}.

\section{Geometrical operators} \label{sec5}
The area and the volume operators can be constructed as in LQG \cite{discr}. The area operator of a surface $S$ can be written as $\hat{A}[S]=\sqrt{\pi_{IJ}(S)\pi^{IJ}(S)}$ and the operator under square root provides just the insertion of the two SU(2) Casimir operators
\begin{eqnarray}
\pi_{IJ}(S)\pi^{IJ}(S) \,D^{j^{(A)}_e}(g^{(A)}_e) \otimes D^{j^{(B)}_e}(g^{(B)}_e)=\\
J^{(A)}_iJ^{(A)}_iD^{j^{(A)}_e}(g^{(A)}_e) \otimes D^{j^{(B)}_e}(g^{(B)}_e) 
 +\,D^{j^{(A)}_e}(g^{(A)}_e) \otimes J^{(B)}_{i}J^{(B)}_{i}D^{j^{(B)}_e}(g^{(B)}_e)\,.\nonumber
\end{eqnarray}
The result is unchanged under $AB$ symmetrization. The final area operator is just the sum of the LQG area operators for each SU(2) group and it is free of Immirzi ambiguity. This is not an issue for the black-hole entropy calculation, since the correct expression can be reproduced in the so-called local perspective \cite{Perez:11}.  

The volume operator can be defined as the square root of the  modulus of the operator $\hat\pi$ corresponding to $\pi$ (\ref{pi}), $\hat{V}[\Sigma]=\sqrt{|\hat{\pi}|}=(\hat{\pi}^2)^{1/4}$. $\hat\pi$ contains three momenta acting at different edges, thus it is nontrivial only at the nodes of the graph. The action of the three-momenta operator $\hat{O}=\epsilon^{IJKL} \pi^{\phantom1 M}_I(S^1)\,\pi_{MJ}(S^2)\,\pi_{KL}(S^3)$ gives 
\begin{eqnarray}
\hat{O}\left(\,D^{j^{(A)}_{e_1}}(g^{(A)}_{e_1}) D^{j^{(A)}_{e_2}}(g^{(A)}_{e_2}) D^{j^{(A)}_{e_3}}(g^{(A)}_{e_3})\otimes D^{j^{(B)}_{e_1}}(g^{(B)}_{e_1})D^{j^{(B)}_{e_2}}(g^{(B)}_{e_2})D^{j^{(B)}_{e_3}}(g^{(B)}_{e_3})\right)=\nonumber\\
\epsilon^{ijk} \,J^{(A)}_iD^{j^{(A)}_{e_1}}(g^{(A)}_{e_1}) J^{(A)}_jD^{j^{(A)}_{e_2}}(g^{(A)}_{e_2}) J^{(A)}_kD^{j^{(A)}_{e_3}}(g^{(A)}_{e_3})
\otimes D^{j^{(B)}_{e_1}}(g^{(B)}_{e_1})D^{j^{(B)}_{e_2}}(g^{(B)}_{e_2})D^{j^{(B)}_{e_3}}(g^{(B)}_{e_3})\nonumber\\
-\epsilon^{ijk} \,D^{j^{(A)}_{e_1}}(g^{(A)}_{e_1}) D^{j^{(A)}_{e_2}}(g^{(A)}_{e_2}) D^{j^{(A)}_{e_3}}(g^{(A)}_{e_3})
\otimes J^{(B)}_iD^{j^{(B)}_{e_1}}(g^{(B)}_{e_1})J^{(B)}_jD^{j^{(B)}_{e_2}}(g^{(B)}_{e_2})J^{(B)}_kD^{j^{(B)}_{e_3}}(g^{(B)}_{e_3})\,,\nonumber
\end{eqnarray}
thus it is the difference of the analogous LQG operator acting on the two SU(2) subgroups. For instance, the Lorentz-invariant state built from the eigenstates of the LQG volume operator with eigenvalues $v_A$ and $v_B$, is an eigenstate of the volume operator with eigenvalue $\sqrt{|v_A^2-v_B^2|}$.      

\section{Scalar constraint} \label{sec6}
The Euclidean part of the scalar constraint in LQG is the product of the connection curvature with two momenta and the inverse square root of the momenta determinant. The expression of the scalar constraint $\mathcal{S}$ in Eq.(\ref{const}) coincides with the Euclidean scalar constraint of LQG, except that the internal group is $O(1,3)$ instead of $SU(2)$. Hence, the scalar constraint $\mathcal{S}$ can be defined using the standard approach \footnote{Other approaches can be equally applied here, as for instance the so-called Warsaw Hamiltonian \cite{Alesci:15}.} based on the so-called Thiemann trick, that allows to rewrite it in terms of the volume operators. Finally, the scalar constraint is quantized as
\begin{equation}   
\hat{\mathcal{S}}\propto \epsilon^{abc}\, Tr(h_{ab} h_c[\sqrt{|\hat{\pi}|}, h_c^{-1}])\,, \label{s}
\end{equation}
where $h_{ab}$ is the holonomy along a loop in the directions $a$ and $b$, while $h_c$ is the holonomy along the direction $c$ (when acting on a quantum states $a$, $b$ and $c$ are adapted to the directions of the graph edges). This expression can be straightforwardly applied here simply taking $h_{ab}$ and  $h_c$ as Lorentz representations of the same kind as those in $\Psi_s$, {\it i.e.} of the form $(j_1,j_2)\oplus (j_2,j_1)$. For instance one could consider the fundamental representation $(1/2,0)\oplus (0,1/2)$. The action of such kind of holonomies is implemented through $SU(2)_A\otimes SU(2)_B$ recoupling theory and it maps Lorentz-invariant spin-networks among themselves. Hence, the full scalar constraint can be computed as two copies of the Euclidean LQG scalar constraint. In other words, the computational complexity of the dynamics is lower than in LQG, since the Lorentzian part is not present.

\section{Conclusions} \label{sec7}
It has been shown how it is possible to apply loop quantization to the manifestly Lorentz-covariant vier-bein formulation of gravity. This analysis has profound implications on LQG and on its interplay with fundamental particles. 
From the point of view of LQG, the formulation is free of the Immirzi ambiguity and the scalar constraint is just the so-called Euclidean part. Concerning the implications in particle physics, it has been elucidated the relationship between the $SU(2)$ group proper of LQG and the Lorentz symmetry, by providing a representation of gravitation quantum states in terms of Lorentz finite representations, whose fundamental elements are Dirac spinors. The same representations have been used by Wigner to classify fundamental particles. Therefore, this work links LQG and Quantum Field Theory, by defining an arena for the implementation of fundamental particle fields in a Quantum Gravity theory with \emph{the relativistic spin as the gravitational charge}.

The proposed model is explicitly symmetric under the exchange of the two $SU(2)$ subgroups, which correspond to the left-handed and right-handed projections of the Lorentz group. In future developments, it would be interesting to discuss how to reconcile this formulation with the chirality of the electro-weak model.



\section*{References}

\begin{thebibliography}{00}


\bibitem{Wigner:39}
E.P. Wigner, Annals of Mathematics, 40(1), (1939) 149.

\bibitem{Bargmann:48}
V. Bargmann and E.P. Wigner, {\it Proc Natl Acad Sci USA} 34(5), (1948) 211.

\bibitem{Weinberg:02}
S. Weinberg, ``The Quantum Theory of Fields, 1'', Cambridge University Press, Cambridge, (2002). 

\bibitem{revloop}
C. Rovelli, ``Quantum gravity'', Cambridge University Press, Cambridge, (2004). 

\bibitem{Thiemannb}
T. Thiemann, ``Modern Canonical Quantum General Relativity'', Cambridge University Press, Cambridge, 2006.

\bibitem{Cianfrani:09}
F. Cianfrani, G. Montani, {\it Phys. Rev. Lett.}, {\bf 102}, (2009) 091301. 

\bibitem{Immirzi}
C. Rovelli, T. Thiemann, {\it Phys.Rev. D}, {\bf 57}, (1998) 1009.

\bibitem{ALMMT95}
A. Ashtekar, J. Lewandowski, D. Marolf, J. Mourao, T. Thiemann, {\it J. Math. Phys.}, {\bf 36}, (1995), 6456-6493. 

\bibitem{LOST06}
J. Lewandowski, A. Okolow, H. Sahlmann, T. Thiemann, {\it Comm. Math. Phys.}, {\bf 267}, No. 3,(2006) 703-733. 

\bibitem{discr}
C. Rovelli, L. Smolin, {\it Nucl. Phys. B}, {\bf 442}, (1995), 593-622.\\
A. Ashtekar, J. Lewandowski, {\it Class. Quant. Grav.}, {\bf 14}, (1997), A55-A82.


\bibitem{Thiemann:98qsd}
T. Thiemann, {\it Class. Quant. Grav.}, {\bf 15}, (1998) 875.

\bibitem{Thiemann:07}
T. Thiemann, {\it Lect. Notes Phys.}, {\bf 721}, (2007) 185.

\bibitem{Ashtekar:86}
A. Ashtekar, {\it Phys. Rev. Lett.}, {\bf 57}, (1986) 2244. 

\bibitem{Barrett:98}
J.W. Barrett, L. Crane, {\it J. Math. Phys.}, {\bf 39}, (1998) 3296.

\bibitem{Barrett:00}
J.W. Barrett, L. Crane, {\it Class. Quant. Grav.}, {\bf 17}, (2000) 3101.

\bibitem{Crane:01}
L. Crane, A. Perez and C. Rovelli, {\it Phys. Rev. Lett.}, {\bf 87}, (2001) 181301.

\bibitem{Thiemann:98}
T. Thiemann, {\it Class. Quant. Grav.}, {\bf 15}, (1998) 1487.

\bibitem{Barbero:95}
J. F. Barbero, {\it Phys. Rev. D}, {\bf 51}, (1995) 5507. 

\bibitem{Cianfrani:14}
F. Cianfrani, O.M. Lecian, M. Lulli, G. Montani, ``Canonical Quantum Gravity: Fundamentals and Recent Developments'',  World Scientific, Singapore, (2014).

\bibitem{Bodendorfer:13}
N. Bodendorfer, T. Thiemann, A. Thurn, {\it Class. Quantum Grav.}, {\bf 30}, (2013) 045001.

\bibitem{Ho96}
S. Holst, {\it Phys. Rev. D}, {\bf 53}, (1996) 5966-5969. 

\bibitem{BS01}
N. Barros e Sa, {\it Int. J. Mod. Phys. D}, {\bf 10}, (2001) 261.

\bibitem{Alexandrov:2006}
S. Alexandrov, E. Buffenoir, P. Roche, {\it Class. Quant. Grav.}, {\bf 24}, (2007) 2809. 

\bibitem{Brink:68}
D.M. Brink, G.R. Satchler, ``Angular momentum'', Clarendon Press, Oxford, (1968).

\bibitem{Makinen:19}
I. M$\ddot{\mathrm{a}}$kinen, ``Introduction to SU(2) Recoupling Theory and Graphical Methods for Loop Quantum Gravity'', arXiv:1910.06821.

\bibitem{Alexandrov:03}
S. Alexandrov and E.R. Livine, {\it Phys.Rev. D}, {\bf 67}, (2003) 044009.

\bibitem{Perez:11}
A. Ghosh and A. Perez, {\it Phys. Rev. Lett.}, {\bf 107}, (2011) 241301. 

\bibitem{Alesci:15} 
E. Alesci, M. Assanioussi, J. Lewandowski and I. Mäkinen, {\it Phys. Rev. D}, {\bf 91}, (2015) 124067.


\end{thebibliography}


\end{document}